\documentclass{article}
\usepackage{spconf,amsmath,graphicx,hyperref}
\usepackage{color}
\usepackage{cite}
\usepackage[T1]{fontenc}
\usepackage{amsmath}
\usepackage{amssymb}
\usepackage{xspace}
\usepackage{hyperref}

\DeclareRobustCommand{\AmbER}{AmbER\textsuperscript{2}\xspace}

\title{AmbER\textsuperscript{2}\xspace: Dual Ambiguity-Aware Emotion Recognition Applied to Speech and text}
\name{Jingyao Wu\textsuperscript{*}\thanks{*Jingyao Wu is funded by the MIT–Novo Nordisk Artificial Intelligence Postdoctoral Fellowship; corresponding author: jingyaow@mit.edu.}, Grace Lin, Yinuo Song, Rosalind Picard}
\address{Massachusetts Institute of Technology}
\begin{document}
\ninept
\maketitle
\begin{abstract}
Emotion recognition is inherently ambiguous, with uncertainty arising both from rater disagreement and from discrepancies across modalities such as speech and text. 
There is a growing focus on modeling rater ambiguity using label distributions. However, the modality ambiguity is not well addressed, and multimodal approaches have focused on feature fusion without explicitly addressing conflicts between modalities. 
In this work, we propose \AmbER, a dual ambiguity-aware framework that simultaneously models rater- and modality-level ambiguity through a teacher–student architecture with a distribution-wise training objective. 
Evaluations on IEMOCAP and MSP-Podcast demonstrate that \AmbER consistently improves distributional fidelity over conventional cross-entropy baselines and achieves performance competitive with or superior to recent state-of-the-art systems. For example, on IEMOCAP, \AmbER achieves relative improvements of $20.3\%$ on Bhattacharyya coefficient (0.83 .vs 0.69), $13.6\%$ on $R^2$ (0.67 .vs 0.59), $3.8\%$ on accuracy (0.683 .vs 0.658), and $4.5\%$ on F1 (0.675 .vs 0.646). Further analysis across ambiguity levels reveals that explicitly modeling ambiguity is particularly beneficial for highly uncertain samples. These findings highlight the importance of jointly addressing rater and modality ambiguity for building robust emotion recognition systems.

\end{abstract}
\begin{keywords}
speech emotion recognition, rater ambiguity, rater uncertainty, multimodal emotion modelling
\end{keywords}
\vspace{-0.5em}
\section{Introduction}
\label{sec:intro}

Speech emotion recognition is a key component in building natural and effective human–machine interactions~\cite{wani2021comprehensive}. However, the inherently ambiguous and subjective nature of emotions poses significant challenges for learning reliable models \cite{sethu2019ambiguous}. Typically, emotion labels are collected from a group of human annotators who provide perceptual evaluations of the same recording. The differences in perception, referred to as inter-rater ambiguity, reflect the subtle and subjective nature of emotions. Recent work has emphasized the need to model ambiguity explicitly, recognizing that disagreement among annotators is informative and should be integrated into affect recognition models to better capture the complexity of emotional understanding \cite{han2017hard, wu2022novel, bose2024continuous, wu2024can, wu24_interspeech}. Rather than collapsing annotations into a single ground-truth label via majority vote or averaging, many appraoches model ambiguity with probability distributions, which has led to the development of ambiguity-aware emotion recognition systems that predict label distributions~\cite{wu2022novel, wu24_interspeech, wu2024emo, chou2024embracing}.  


In contrast, another source of ambiguity has received comparatively little attention: modality ambiguity. 
This arises when different modalities convey conflicting emotional cues—for example, prosodic features of speech may suggest anger while the lexical content of the same utterance indicates neutrality. 
Beyond such sample-level conflicts, modality ambiguity also reflects systematic differences in how modalities capture emotion, which in turn affects emotion recognition performance for different modalities \cite{goncalves2025jointly}. 
Prior studies have shown, for instance, that arousal is more reliably conveyed through acoustic features while valence is often better captured through facial expressions in video \cite{schoneveld2021leveraging}. Both rater- and modality-level ambiguity exist and play important roles in emotion modeling; however, no existing framework has simultaneously modeled the two sources of ambiguity.

In this paper, we propose the novel Dual Ambiguity-Aware Emotion Recognition framework, \AmbER{}, addressing two sources of ambiguity in this work: \textit{rater ambiguity} and \textit{modality ambiguity}. 
\AmbER adopts a teacher–student architecture with a novel Ambiguity-Integrated training objective, in which modality-specific heads act as experts and provide adaptive guidance to a student head, while the student is simultaneously supervised by ground-truth label distributions from human annotators. 
Experimental results on IEMOCAP and MSP-Podcast across a range of evaluation metrics show that \AmbER{} consistently outperforms conventional cross-entropy baselines and achieves state-of-the-art performance. Moreover, our analysis across different ambiguity levels reveals that explicitly modeling ambiguity is especially beneficial for highly ambiguous samples. 

\begin{figure*}[t!]
    \centering
    \includegraphics[width=0.85\textwidth, height=0.14\textheight]{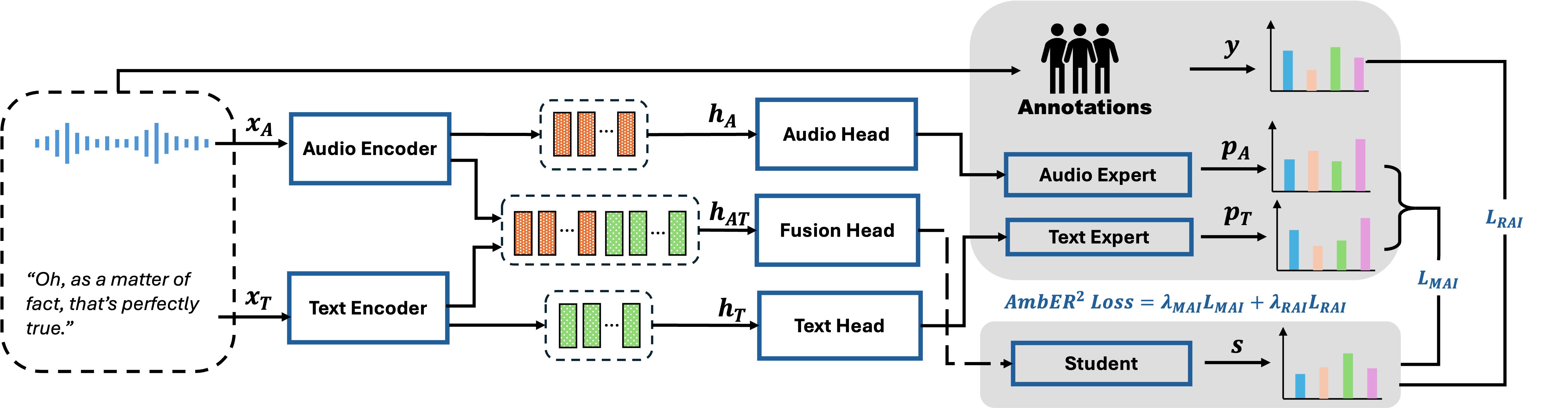}\vspace{-0.5em} 
    \caption{An graphical illustration of the proposed \AmbER{} framework when \textit{Fusion Head} is considered as \textit{Student}.}
    \label{fig:amber}\vspace{-2em}
\end{figure*}

\vspace{-1em}
\section{Related Work}

Current approaches to ambiguity modeling move beyond single “ground truth” labels toward distributional representations that capture variability across annotators. For categorical emotion tasks, soft-label distributions represent the proportion of annotations for each class \cite{chou2024embracing, fayek2016modeling, lotfian2017formulating, hong2025aer}, while multi-label classification allows multiple emotions to be assigned to a single instance, reflecting co-occurring or blended states \cite{ju2020transformer, ando2019speech}. More recently, \cite{chou2024embracing} proposed an “all-inclusive rule” that incorporates all ratings and distributional labels as multi-label targets during training. For continuous arousal–valence prediction, emotional states have been modeled as probability distributions (e.g., Gaussian \cite{han2017hard}, Beta \cite{bose2024continuous}, or non-parametric \cite{wu2022novel} forms) to better capture rater ambiguity. However, these efforts have primarily focused on rater ambiguity, leaving modality ambiguity largely unaddressed. 

There have been great efforts in building multimodal emotion recognition systems, benefiting from advances in fusion strategies \cite{wu2014survey} such as early fusion (merging feature embeddings) \cite{ poria2018multimodal}, intermediate fusion (joint learning in a shared embedding space)\cite{chen2024modality}, and late fusion (combining decisions at the output level)\cite{georgescu2024exploring}. While these approaches aim to improve performance, they do not explicitly address conflicts between modalities. Existing works have examined how modalities align with specific emotion classes or affect dimensions, but such modality ambiguity is not directly modeled in system design \cite{wu2021multimodal, tzirakis2019real}. More recently, \cite{chua2025speech} proposed an entropy-aware score selection method to combine speech and text predictions. Nevertheless, modality ambiguity has never been jointly investigated with rater ambiguity, leaving a critical gap in developing robust, dual ambiguity-aware emotion recognition frameworks.

\vspace{-1em}
\section{Proposed \AmbER Framework}

We introduce \AmbER, a framework that explicitly models both rater and modality ambiguity for emotion recognition. The design follows a teacher–student paradigm, where modality-specific heads act as experts guiding a designated student head. As shown in Fig.~\ref{fig:amber}, the audio and text heads serve as experts, while the fusion head is the student (though roles are interchangeable). All networks are trained jointly: the student receives adaptive supervision from the modality experts ($p_A$, $p_T$) and is simultaneously supervised by the ground-truth soft label distributions $y$, preserving both rater and modality-level ambiguity.


\vspace{-1em}
\subsection{Model Structure}
Assume the emotion labels collected from $N$ human annotators for a given sample are denoted as $\{l_1, l_2, \dots, l_N\}$, where each $l_i \in \{1, \dots, C\}$ corresponds to one of $C$ emotion classes. A soft label distribution $y \in \mathbb{R}^C$ can then be obtained by aggregating these annotations by normalizing vote counts into a probability distribution, which will serves as supervision during training: 

\begin{equation}
    y_c = \frac{n_c}{N}, \quad c = 1, \dots, C, \quad \sum y_c = 1
\end{equation}
where $n_c$ denote the number of annotators assign class $c$.

Modality-specific outputs are obtained through each modality head as defined in Eq. \ref{eq:m}.

\vspace{-0.5em}
\begin{equation}\label{eq:m}
    p_m = g_m(h_m), \quad m \in \{A, T, AT\},
\end{equation}
where $h_m$ denotes hidden representations extracted from modality specific encoders given paired audio and text inputs, denoted $x_A$ and $x_T$.
$g_m(\cdot)$ represents a network that maps hidden representations into a probability distribution over emotion classes.  


Among these predictions, one head is designated as the \emph{student}, denoted $s = p_{m^*}$, while the remaining heads act as \emph{experts}. 
Depending on the configuration, the student may be unimodal (A or T) or multimodal (AT), with the corresponding experts providing complementary guidance. 
For the sake of comprehensive performance, this paper adopts the structure where the $Fusion$ head serves as the $student$, as it can integrate information across modalities while distilling complementary signals from unimodal experts and aligning with rater distributional ambiguity.

\vspace{-1em}
\subsection{Dual Ambiguity Training}

We define two complementary loss terms:  
(i) the \emph{Rater Ambiguity Integrated (RAI) Loss}, which enforces fidelity to rater distributions, and  
(ii) the \emph{Modality Ambiguity Integrated (MAI) Loss}, which adaptively aligns the student with modality experts. 
Together, these form the overall \AmbER{} loss.  

Specifically, to preserve rater ambiguity, the student prediction $s$ is directly matched to the ground-truth distribution $y$:
\vspace{-0.5em}
\begin{equation}
    \mathcal{L}_{RAI} = \mathrm{JS}(y \parallel s),
\end{equation}
where $\mathrm{JS}(\cdot \parallel \cdot)$ denotes Jensen–Shannon divergence.  

To address modality ambiguity, the key idea is that not all experts should contribute equally to the student. 
When experts are closer to the ground truth, their predictions are more reliable; when they deviate from it, their supervision should be down-weighted. 

Let $\mathcal{M} = \{A, T, AT\}$ denote the set of modality-specific heads, where $m^* \in \mathcal{M}$ is the index of the designated student and the remaining heads ($m \in \mathcal{M}, m \neq m^*$) act as experts. 
For example, if the fusion head (AT) is chosen as the student, then the audio (A) and text (T) heads serve as experts.  

We therefore introduce a weighted consistency loss, where the contribution of each expert is scaled by its agreement with the ground truth:
\vspace{-1.2em}
\begin{equation}
    \mathcal{L}_{MAI} = \sum_{m \neq m^*} u_m \, \mathrm{JS}(s \parallel p_m).
\end{equation}
The adaptive weights are defined as
\vspace{-0.5em}
\begin{equation}\label{eq:um}
    u_m = \frac{\exp(-\kappa D_m)}{\sum_{m' \neq m^*} \exp(-\kappa D_{m'})},
    \qquad
    D_m = \mathrm{JS}(p_m \parallel y),
\end{equation}
where $\kappa$ is the sharpness parameter.

Therefore, the experts whose predictions better match the rater distribution (small $D_m$) receive larger weights.  
In this formulation, unreliable experts are down-weighted, while more rater-consistent experts exert stronger influence. 
This ensures that the student learns primarily from the most reliably modality signals, while still accounting for variability across modalities.

Finally, the overall \AmbER{} loss integrates both terms:
\begin{equation}
    \mathcal{L}_{\text{AmbER}} = \lambda_{RAI} \, \mathcal{L}_{RAI} + \lambda_{MAI} \, \mathcal{L}_{MAI},
\end{equation}
where $\lambda_{RAI}$ and $\lambda_{MAI}$ balance the contributions of rater- and modality-level ambiguity.

\vspace{-0.5em}
\section{Experimental Settings}
\vspace{-1em}
\subsection{Datasets}

\noindent\textbf{IEMOCAP.} \; The Interactive Emotional Dyadic Motion Capture (IEMOCAP) \cite{busso2008iemocap} database contains approximately 12 hours of audiovisual recordings of scripted and improvised dyadic interactions, seprated by five sessions. Each utterance is approximately 4.5 second length, annotated by multiple raters with categorical emotion labels. Following standard practice as in \cite{hong2025aer, wu2024emo}, we adopt the 4-class setting comprising \{\emph{Neutral}, \emph{Happy}, \emph{Angry}, and \emph{Sad}\}. 

\noindent\textbf{MSP-Podcast.} \; The MSP-Podcast corpus (v1.12) \cite{lotfian2017building} consists of natural, unscripted speech segments collected from podcast recordings. Each segment is 2-10 second lengths, annotated by at least five raters, yielding categorical labels as well as distributional information that captures inter-rater variability. We follow prior work and use the 8-class setting \cite{hong2025aer}, which includes emotions such as \{\emph{Neutral}, \emph{Happy}, \emph{Sad}, \emph{Angry}, \emph{Disgust}, \emph{Fear}, \emph{Surprise}, and \emph{Contempt}\}. 

\vspace{-1em}
\subsection{Implementations}
\noindent\textbf{Feature Encoders.} Wav2Vec2 \cite{baevski2020wav2vec} pre-trained encoder is employed for audio feature extraction, which has demonstrated strong performance in emotion recognition tasks \cite{chua2025speech, wu2024emo}. We extract frame-level embeddings and mean pooling across time to obtain utterance-level representations. For text, we adopt the pre-trained BERT model \cite{devlin2019bert}, and obtain sentence-level representations by mean-pooling the last hidden states weighted by the attention mask. For multimodal fusion, the audio and text embeddings are concatenated and combined via a gated fusion mechanism to form the fused representation, which is then passed to the fusion head to produce predictions over the emotion classes.





\noindent\textbf{Baseline.} 
As a baseline without ambiguity modeling, we adopt the same architecture as \AmbER{} but train it with Class-Balanced Cross-Entropy (CB-CE) loss, following prior work for emotion recognition with soft distributions \cite{wu2024emo}.  

\noindent\textbf{Network Parameters.} 
Each head (A, T, AT) is implemented as a two-layer MLP, consisting of two linear layers with ReLU activation and a final softmax for distributional outputs. All heads are trained jointly with the proposed \AmbER{} loss, where one head is designated as the student and the remaining heads act as experts. The balancing coefficients are set to $\lambda_{\text{RAI}}=1.0$ and $\lambda_{\text{MAI}}\in\{0.3,0.5,0.7\}$, selected via validation, and the sharpness parameter $\kappa$ in Eq.~\ref{eq:um} is tuned over $\{2,4,8\}$.


\noindent\textbf{Training.} 
We train with AdamW (learning rate $3\times10^{-4}$, weight decay $10^{-2}$), batch size 128, and a maximum of 30 epochs. Five-fold cross-validation is applied to both datasets. In IEMOCAP, each session serves as a fold. For MSP-Podcast, we use five equal-sized folds rather than the standard speaker partitions to better balance ambiguous samples across splits. Each experiment is repeated with five random seeds, and both the mean and standard deviation of results are reported. All implementations are based on PyTorch, and hyperparameters are selected based on validation performance.




\noindent\textbf{Evaluation Metrics.} 
Distribution-based metric are employed follow \cite{hong2025aer}, including Jensen–Shannon divergence (JS), Bhattacharyya coefficient (BC), and $R^2$. All three metrics are bounded in $[0,1]$: lower JS indicates a closer match between predicted and ground-truth distributions, higher BC reflects greater similarity, and higher $R^2$ reflects a better goodness-of-fit. To ensure comparability with conventional emotion recognition studies, we additionally report classification-based metrics, including macro-F1, weighted-F1 (W-F1), and accuracy (ACC). It is noted that these metrics are less comparable since they do not account for ambiguity.   

\vspace{-1em}
\section{Results}
\vspace{-0.5em}

\subsection{Baseline without Ambiguity Modelling}
Table \ref{tab:CE_js} and \ref{tab:CE_F1} present the baseline performance when models are trained with conventional CB-CE loss, i.e., without explicitly modeling ambiguity.  On both IEMOCAP and MSP-Podcast, the fusion model ($p_{AT}$) achieves the strongest performance across all three metrics, consistently outperforming unimodal systems ($p_T$, $p_T$). This confirms the complementary nature of audio and text modalities in emotion recognition. A closer inspection further reveals modality-specific differences: for example, on IEMOCAP, the audio model ($p_A$) performs better than the text model ($p_T$) on most metrics (e.g., JS of 0.275 vs.\ 0.302, BC of 0.747 vs.\ 0.723). Such discrepancies could arise from uncertainties in the encoders themselves and their capacity to capture emotion-relevant cues, but may also partly reflect modality-dependent ambiguity. Nevertheless, these differences highlight the uneven ability of individual modalities to decode ambiguity-aware emotion states, which motivates our subsequent analysis with the proposed \AmbER framework.

\begin{table}[tb!]
\vspace{-1em}
\centering
\small
\caption{Performance of baseline systems for different modalities when trained with Class-Balanced Cross Entropy loss, evaluated with distribution based metrics.}
\label{tab:CE_js}
\resizebox{\columnwidth}{!}{
\begin{tabular}{lccc}
\hline
\multicolumn{4}{c}{\textbf{IEMOCAP}} \\
\hline
\textbf{Modality} & \textbf{JS} $\downarrow$ & \textbf{BC} $\uparrow$ & \textbf{R$^2$} $\uparrow$ \\
\hline
Text ($P_t$)   & $0.302 \pm 0.001$ & $0.723 \pm 0.001$ & $ 0.540 \pm 0.001$          \\
Audio ($P_a$)   & $0.275 \pm 0.004$ & $0.747 \pm 0.003$ & $0.526 \pm 0.003$   \\
Audio + Text ($P_{a+t}$) & $\boldsymbol{0.216 \pm 0.001}$ & $\boldsymbol{0.803 \pm 0.001}$ & $\boldsymbol{0.628 \pm 0.001}$   \\
\hline
\multicolumn{4}{c}{\textbf{MSP-Podcast}} \\
\hline
\textbf{Modality} & \textbf{JS} $\downarrow$ & \textbf{BC} $\uparrow$ & \textbf{R$^2$} $\uparrow$ \\
\hline
Text ($P_t$) & $ 0.386\pm 0.001$ & $ 0.648 \pm 0.001$ & $0.355\pm 0.005$             \\
Audio ($P_a$)  & $0.388\pm 0.001$ & $ 0.646 \pm 0.001$ & $0.359 \pm 0.002$          \\
Audio + Text ($P_{a+t}$) & $\boldsymbol{0.368\pm0.003}$ & $\boldsymbol{0.664 \pm 0.000}$ & $\boldsymbol{0.378 \pm 0.002}$  \\
\hline
\end{tabular}
}\vspace{-1.5em}
\end{table}

\begin{table}[!tb]
\centering
\caption{Performance of baseline systems for different modalities when trained with Class-Balanced Cross Entropy loss, evaluated with conventional classification metrics.}
\label{tab:CE_F1}
\resizebox{\columnwidth}{!}{%
\begin{tabular}{lccc}
\hline
\multicolumn{4}{c}{\textbf{IEMOCAP}} \\
\hline
\textbf{Modality} & \textbf{F1} $\uparrow$ & \textbf{W-F1} $\uparrow$ & \textbf{ACC} $\uparrow$ \\
\hline
Text ($P_t$)  & $0.581 \pm 0.007$ & $0.574\pm 0.005$ & $ 0.571 \pm 0.006$ \\
Audio ($P_a$) & $0.654 \pm 0.004$ & $0.538 \pm 0.006$ & $ 0.544 \pm 0.006$  \\
Audio + Text ($P_{a+t}$) & $\boldsymbol{0.690 \pm 0.003}$ & $\boldsymbol{0.655 \pm 0.003}$ & $\boldsymbol{0.654 \pm 0.003}$ \\
\hline
\multicolumn{4}{c}{\textbf{MSP-Podcast}} \\
\hline
\textbf{Modality} & \textbf{F1} $\uparrow$ & \textbf{W-F1} $\uparrow$ & \textbf{ACC} $\uparrow$ \\
\hline
Text ($P_t$)   &$0.247 \pm 0.003$ & $ 0.535 \pm 0.003$ & $0.446 \pm 0.007$ \\
Audio ($P_a$)    &$0.247 \pm 0.001$ & $0.522 \pm 0.001$ & $\boldsymbol{0.478 \pm 0.002}$ \\
Audio + Text ($P_{a+t}$) & $\boldsymbol{0.276\pm0.001}$ & $\boldsymbol{0.552 \pm 0.002}$ & $0.473 \pm 0.003$ \\ 
\hline
\end{tabular}%
}\vspace{-1.5em}
\end{table}

\vspace{-1em}
\subsection{Modelling with \AmbER{}}
Tables~\ref{tab:at_jsbc} 
summarizes the performance of our proposed \AmbER{} framework when applied to audio and text embeddings. Compared to the baseline trained with CB-CE loss, \AmbER{} consistently improves distribution-based metrics on both IEMOCAP and MSP-Podcast. For instance, on IEMOCAP, \AmbER{} reduces JS from $0.216$ to $0.193$ (relative decrease 10.6\%$\downarrow$), improves BC from $0.803$ to $0.825$ (2.7\%$\uparrow$), and achieves higher $R^2$ (5.9\%$\uparrow$). Similarly, on MSP-Podcast, JS decreases from $0.368$ to $0.328$ (10.9\%$\downarrow$), BC improves from $0.664$ to $0.707$ (6.5\%$\uparrow$), while $R^2$ rises from $0.378$ to $0.425$ (12.4\%$\uparrow$). These results demonstrate that explicitly training with rater and modality ambiguity leads to more reliable distributional predictions.



On conventional classification metrics reported in Table~\ref{tab:at_cls}, \AmbER{} also delivers consistent gains on IEMOCAP with improving W-F1 
(3.1\%$\uparrow$) and accuracy 
(4.4\%$\uparrow$). On MSP-Podcast, accuracy rises from $0.473$ to $0.520$ (9.9\%$\uparrow$), although  W-F1 shows a slight decline. This trade-off reflects the challenge of simultaneously optimizing distributional fidelity and hard single-class decision boundaries. Overall, the improvements across classification metrics, together with the stronger distribution-based results, confirm the effectiveness of explicitly modelling the dual ambiguity (rater and modality) in the proposed \AmbER{} framework.

\vspace{-1em}
\subsection{Comparison with State of the Art}

We compare our proposed \AmbER{} framework with prior state-of-the-art systems in Table~\ref{tab:stoa}. 
We consider three representative models. AER-LLM (Gemini-1.5-Flash, 2025) serves as the closest comparison, as it also leverages speech and text modalities and reports the same evaluation metrics; we include both its zero-shot (ZS) and few-shot (FS) settings \cite{hong2025aer}. For conventional single-class evaluation, we compare against a recent method, denoted as $\text{Emo}_\text{ent}$ (2025), which incorporates entropy-aware training with speech and text modalities \cite{chua2025speech}. 
Finally, we include EMO-Super (2024), a benchmark that evaluates a range of self-supervised speech representations \cite{wu2024emo}. 
We report their results using the Wav2Vec2 setting, which achieves strong performance on IEMOCAP and competitive results on MSP-Podcast, though it relies solely on the speech modality.

On IEMOCAP, \AmbER{} achieves consistent improvements across both distributional and classification evaluations, outperforming all prior systems and even surpassing the few-shot setting of AER-LLM, with $45.7\%$ relative improvement on JS, $20.3\%$ on BC, and $13.6\%$ on $R^2$. 
On MSP-Podcast, \AmbER{} similarly improves distributional metrics, with the highest particularly JS and BC, while maintaining competitive $R^2$. 
Although W-F1 score is lower due to the increased difficulty of distributional supervision on this naturalistic dataset, particularly pronounced for ambiguous minority classes, \AmbER{} still exceeds prior work in accuracy under the zero-shot setting and demonstrates clear advantages in modeling distributional fidelity. 
Overall, these results confirm that explicitly incorporating rater and modality ambiguity yields more faithful distributional predictions and performance that is competitive with or superior to state-of-the-art systems.

\begin{table}[!tb]
\vspace{-1em}
\centering
\scriptsize
\caption{Performance of \AmbER evaluated with distribution based metrics.}\vspace{-0.5em}
\label{tab:at_jsbc}
\begin{tabular}{lcc}
\hline
\multicolumn{3}{c}{\textbf{IEMOCAP}} \\
\hline
\textbf{Metric} & \textbf{Baseline (A+T)} & \textbf{\AmbER{}} \\
\hline
JS $\downarrow$   & $0.216 \pm 0.001$ & $\boldsymbol{0.193 \pm 0.002}$ \\
BC $\uparrow$ & $0.803 \pm 0.001$ & $\boldsymbol{0.825 \pm 0.001}$ \\
$R^2$ $\uparrow$      & $0.628 \pm 0.001$ & $\boldsymbol{0.665 \pm 0.002}$ \\
\hline
\multicolumn{3}{c}{\textbf{MSP-Podcast}} \\
\hline
\textbf{Metric} & \textbf{Baseline (A+T)} & \textbf{\AmbER{}} \\
\hline
JS $\downarrow$  & $0.368 \pm 0.003$ & $\boldsymbol{0.328 \pm 0.001}$ \\
BC $\uparrow$ & $0.664 \pm 0.000$ & $\boldsymbol{0.707 \pm 0.000}$ \\
$R^2$ $\uparrow$      & $0.378 \pm 0.002$ & $\boldsymbol{0.425 \pm 0.001}$ \\
\hline
\end{tabular}\vspace{-2em}
\end{table}

\vspace{-1em}
\begin{table}[!tb]
\centering
\scriptsize
\caption{Performance of \AmbER evaluated with classification based metrics.}
\label{tab:at_cls}
\begin{tabular}{lcc}
\hline
\multicolumn{3}{c}{\textbf{IEMOCAP}} \\
\hline
\textbf{Metric} & \textbf{Baseline (A+T)} & \textbf{\AmbER{}} \\
\hline
F1 $\uparrow$    & $0.690 \pm 0.003$ & $\boldsymbol{0.695 \pm 0.005}$ \\
W-F1 $\uparrow$ & $0.655 \pm 0.003$ & $\boldsymbol{0.675 \pm 0.004}$ \\
ACC  $\uparrow$          & $0.654 \pm 0.003$ & $\boldsymbol{0.683 \pm 0.003}$ \\
\hline
\multicolumn{3}{c}{\textbf{MSP-Podcast}} \\
\hline
\textbf{Metric} & \textbf{Baseline (A+T)} & \textbf{\AmbER{}} \\
\hline
F1 $\uparrow$    & $0.276\pm0.001$                & $\boldsymbol{0.369 \pm 0.003}$ \\
W-F1 $\uparrow$  & $\boldsymbol{0.552 \pm 0.002}$ & $0.445 \pm 0.002$ \\
ACC  $\uparrow$     & $0.473 \pm 0.003$ & $\boldsymbol{0.520 \pm 0.001}$ \\
\hline
\end{tabular}\vspace{-2em}
\end{table}

\vspace{-1em}
\begin{table}[tb!]
\centering
\caption{Comparison with state-of-the-art.}
\resizebox{\columnwidth}{!}{
\begin{tabular}{lcccccc}
\hline
\multicolumn{7}{c}{\textbf{IEMOCAP}} \\
\hline
\textbf{Method} & \textbf{JS $\downarrow$} & \textbf{BC $\uparrow$} & \textbf{R-Squared $\uparrow$} & \textbf{ACC $\uparrow$} & \textbf{F1 $\uparrow$} & \textbf{W-F1 $\uparrow$} \\
\hline
AER - LLM (2025) \cite{hong2025aer} (ZS) & 0.47 & 0.51 & 0.51 & 0.434 & - & 0.429 \\ 
AER - LLM (2025) \cite{hong2025aer} (FS) & 0.35 & 0.69 &0.59 & 0.481 &-&0.492  \\ 
EMO-Super (2024) \cite{wu2024emo} & - & - & - & - & 0.339 & - \\
$\text{Emo}_\text{ent}$ (2025) \cite{chua2025speech} & - & - & - & 0.658 & 0.646 & - \\
\AmbER (Ours) & \textbf{0.19} & \textbf{0.83} & \textbf{0.67} & \textbf{0.683} & \textbf{0.675} & \textbf{0.675} \\
\hline
\multicolumn{7}{c}{\textbf{MSP-Podcast}} \\
\hline
\textbf{Method} & \textbf{JS $\downarrow$} & \textbf{BC $\uparrow$} & \textbf{R-Squared $\uparrow$} & \textbf{ACC $\uparrow$} & \textbf{F1 $\uparrow$} & \textbf{W-F1 $\uparrow$} \\
\hline
AER - LLM (2025) \cite{hong2025aer} (ZS) & 0.45 & 0.54 & 0.52 & 0.506 & - & 0.505 \\
AER - LLM (2025)\cite{hong2025aer} (FS) & 0.40 & 0.61 & \textbf{0.56} & \textbf{0.556} & - & \textbf{0.562} \\
EMO-Super (2024)\cite{wu2024emo} &- &- &- &- & 0.363 & -\\
\AmbER (Ours) & \textbf{0.33} & \textbf{0.71} & 0.43 & {0.520} & \textbf{0.369} & 0.445 \\

\hline
\end{tabular}\label{tab:stoa}
}\vspace{-1.5em}
\end{table}

\vspace{1em}
\subsection{Analysis from different ambiguity levels}
\begin{figure}[t!]
\includegraphics[width=0.49\textwidth,height=0.2\textheight]{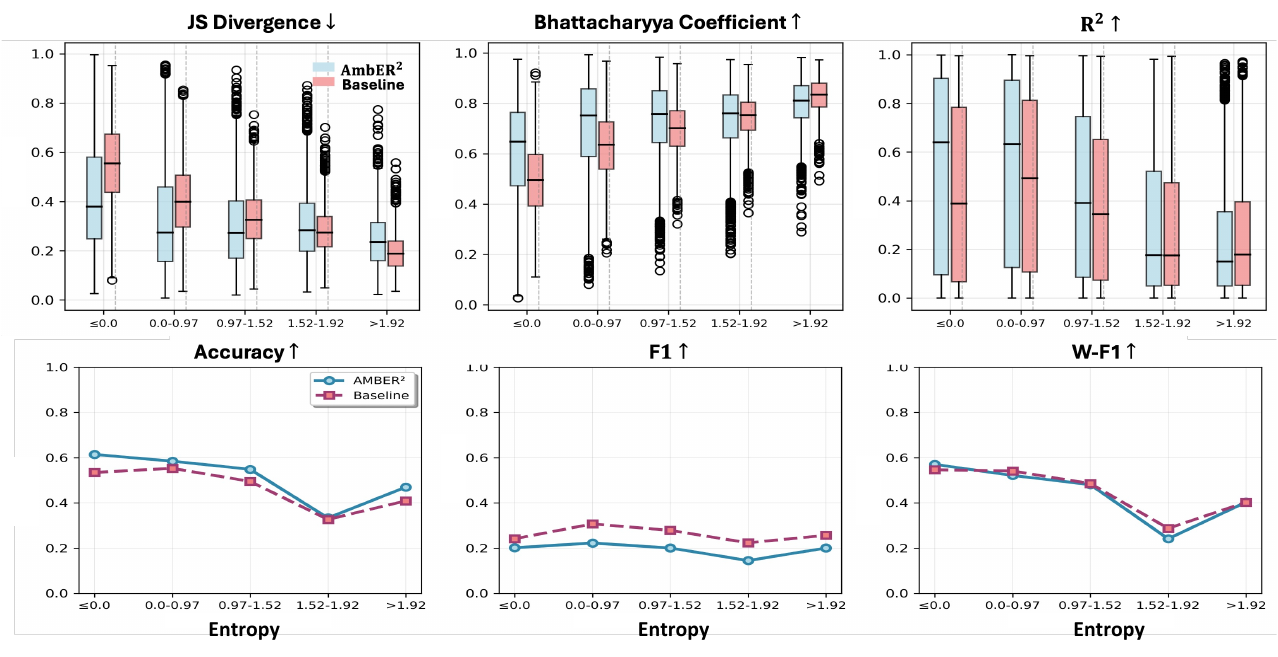}\vspace{-1.5em} 
    \caption{Performance comparison of \AmbER and baseline at different ambiguity levels. }\vspace{-1.5em}
    \label{fig:am_f1}
\end{figure}

We also conducted the in-depth analysis of the proposed \AmbER{} performance at different ambiguity levels follow the similar ideas in \cite{hong2025aer,wu2022novel}. Where entropy of soft distributions aggregated from ground truth annotators are defined as the ambiguity, higher entropy indicate higher ambiguity. This analysis is performed on the MSP-Podcast dataset, where the large number of annotators provides a broader range of ambiguity.

As shown in Figure~\ref{fig:am_f1}(top), \AmbER{} consistently achieves better performances for all evaluation metrics compared to the CB-CE baseline across nearly all ambiguity levels. The improvements are especially pronounced in the low-to-moderate ambiguity ranges (the first two bins), suggesting that explicitly modelling ambiguity helps the model better match the soft label distributions in cases where rater disagreement is non-trivial but not extreme. Interestingly, both the CB-CE baseline and \AmbER{} exhibit decreasing JS and increasing BC as ambiguity rises. This occurs because high-entropy targets are closer to uniform, and the distance between two “fuzzy” distributions is inherently smaller than between sharp ones. However, this does not mean the prediction task itself becomes easier—on the contrary, $R^2$ steadily declines with ambiguity, reflecting the reduced explainable variance and the growing difficulty of capturing fine-grained distributional structure when annotators diverge strongly.


We further evaluate the single-class performance across ambiguity levels (Fig.~\ref{fig:am_f1} bottom), inspired by prior work~\cite{wu2022novel}, which suggests that the single labels (majority or mean) are only representative for low ambiguous samples, whereas high-ambiguity samples should be characterized by full label distributions. It is observed that ACC and W-F1 are highest on low-ambiguity samples, consistent with the fact that clear consensus among raters yields more reliable categorical decisions. As ambiguity increases, performance on these metrics steadily declines for both baseline and \AmbER{}, reflecting the intrinsic difficulty of committing to a single class when annotators themselves diverge, which shows the consistent trend for continuous arousal/valence performances reported by \cite{wu2022novel}. Nevertheless, \AmbER{} generally tracks or slightly outperforms CE at most ambiguity levels, reaffirming that modelling rater distributions does not compromise single-class decision quality, even as the task becomes more challenging in the high-ambiguity regime.


\vspace{-0.5em}
\section{Conclusion}
\vspace{-0.5em}
In this work, we introduced \AmbER{}, a Dual Ambiguity-Aware Emotion Recognition framework that jointly models rater and modality ambiguity through a teacher–student formulation. By combining distributional supervision from human raters with adaptive consistency across speech and text, \AmbER{} produces predictions that better reflect the inherent ambiguity of emotion labels. Experiments on IEMOCAP and MSP-Podcast show that \AmbER{} outperforms strong baselines and achieves competitive results with state-of-the-art systems. Analysis across different ambiguity levels further reveals that explicitly learning from both rater and modality ambiguity is especially beneficial for highly uncertain samples, where conventional methods struggle. These findings highlight the value of treating ambiguity as informative signal, paving the way toward robust and human-aligned affective computing systems. Finally, the framework is generalizable across modalities and languages, and future work will apply it to multimodal emotion recognition and multilingual, cross-cultural datasets, enabling a deeper examination of ambiguity and variability across diverse annotation contexts.

\vfill\pagebreak

\begin{small}
\bibliographystyle{IEEEbib}
\bibliography{strings,refs}
\end{small}

\end{document}